# An Audit Logic for Accountability (Extended Version)[*]

J.G. Cederquist[1], R. Corin[1], M.A.C. Dekker[1,2], S. Etalle[1] and J.I. den Hartog[1]

[1] Department of Computer Science, University of Twente, The Netherlands
{cederquistj,corin,etalle,hartogji}@cs.utwente.nl

[2] Security Department, TNO ICT, The Netherlands
m.a.c.dekker@telecom.tno.nl

**Abstract**

*We describe a policy language and implement its associated proof checking system. In our system, agents can distribute data along with usage policies in a decentralized architecture. Our language supports the specification of conditions and obligations, and also the possibility to refine policies. In our framework, the compliance with usage policies is not actively enforced. However, agents are accountable for their actions, and may be audited by an authority requiring justifications.*

## 1 Introduction

In many situations, there is a need to share data between potentially untrusted parties while ensuring the data is used according to given *policies*. This problem is addressed by two main research streams: on one hand, there is a large body of literature on *access (and usage) control* [8, 16, 11, 4], on the other hand we find *digital rights management* [18, 5]. While the former assumes a trusted *access control service* restricting data access, the latter assume trusted devices in charge of content rendering. Both settings need the trusted components to be available at the moment the request happens, to regulate the data access.

However, there are scenarios (like the protection of private data) in which both access control and digital rights management fail, either because the necessary trusted components are not available or because they are controlled by agents we do not want to trust. For instance, P3P [17] and E-P3P (and also EPAL) [3] are languages that allow one to specify policies for privacy protection; however, the user can only hope that the private data host follows them.

In this paper, the process of regulating the data access is not assumed to be always performed by the same entity at the same moment in which the access is requested. More specifically, we relax this in mainly two ways:

- Firstly, at the moment that the data is requested, we assume that access is always granted, and only later it is determined whether the requestor had permission to access the data. This is the process of *auditing*. To achieve this, we need all the relevant decision information to be kept until audit time (e.g. keeping secure logs).

- Secondly, the entity that is performing the auditing does not need to be fixed, and can thus be dynamically chosen. This is useful since, for example, some authorities are more appropiate to audit specific agents than others. The actual authority does not even need to be one single entity, and can be for example composed of regular agents.

We present a flexible system which allows to express, deploy and reason about policies controlling the usage of data. In our target setting agents can distribute data along with usage policies within a highly decentralized architecture, in which the enforcement of policies is difficult (if not impossible). Therefore, we use instead an *auditing system* with best-effort checking by an authority which is able to observe (some) actions. We introduce a notion of agent *accountability* and express the *proof obligation* of an agent being audited. The system allows to reason about policies and user accountability. Our framework is depicted in Figure 1.

---

[*]This is an extended version of the article to appear in the Conference Proceedings of the 6th IEEE International Workshop on Policies for Distributed Systems and Networks (POLICY 2005)

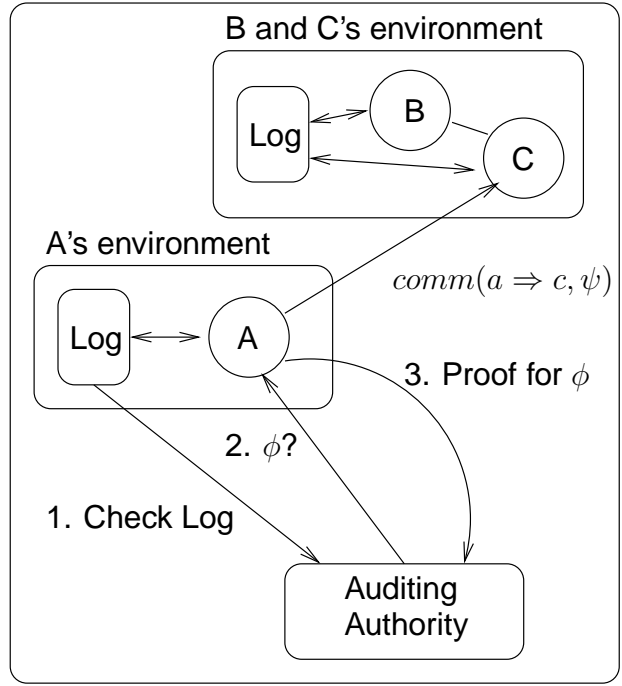

**Figure 1. Our framework**

We make no assumptions on the existence of trusted components regulating access (although we do require a trusted environment to *certify* environmental conditions, and to securely log events). In fact, agents are not forced to follow the policies, but may be audited by authorities which ask for justifications. We make no particular assumptions about authorities; they may comprise, for instance, of groups of regular agents. The more an authority can observe, the more accurate the auditing process is, thus providing more confidence over the agent's behaviour. To characterize compliant agent behaviour, as perceived by an authority, we define *accountability* tests, which are carried out during auditing by the authority.

Of course, our approach does not allow a strict policy enforcement: agents can easily "misbehave" (i.e. treat data in a way that is not allowed by the policy), at risk of being traced. It is our belief that in many emerging scenarios active policy enforcement is infeasible.

This paper builds on the preliminary work reported in [6]. In particular, we provide several extensions, the most notable of which are:

- We include the ability to specify conditions and obligations within the policies.

- Policies may now contain variables and quantifiers. This allows us to define a fundamental rule that gives the ability to refine policies. Agents can create (by refinement) new policies from existing ones, before passing them to other agents. In contrast, in [6] the only policies allowed are those that are explicitly stated by the data owner.

- We precisely describe our system by introducing three functions, namely the *observability*, *conclusion derivation* and *proof obligation* functions. Moreover, we provide a customizable action set to account for particular, user-defined scenarios.

- We define agent accountability tests, and present a (terminating) procedure for recursive auditing.

- Finally, we provide a formalization of our proof system in the proof checker Twelf[13], which allows us to model proofs provided by agents, and the subsequent checking by the authority (Our formalization covers the lower part of Figure 1.)

## 2 A System of Policies and Actions

Our setup consists of a group of agents executing different actions. The permission to execute an action is expressed by a policy constructed using a special logic, introduced below. In this section we introduce some necessary components for our system.

### 2.1 The basics

*Agents* are modelled by a set $\mathcal{G}$ ranged over by $a, b$ and $c$ (referred to as Alice, Bob, and Charlie). We also have a set of agent variables $\mathcal{V}_a$ and use $A, B, C$ to range over both agents and agent variables. Similarly we have a set $\mathcal{D}$ of *data objects*, ranged over by $d$, and a set of data variables $\mathcal{V}_d$. We use $D$ to range over data objects and data variables and $x, y, z$ to range over (data and agent) variables.

Basic *permissions* and *facts* are expressed by atomic predicates in a set $\mathcal{C}$, ranged over by $p$. Examples are read$(a, d)$, which expresses that agent $a$ has permission to read data $d$ and partner$(a, b)$ indicating (the fact) that agent $a$ and $b$ are partners. In general, predicates can relate any number of data objects and agents.

The actions that agents execute are modelled using a set of actions $ACT$, ranged over by $act$. We assume that two types of actions are always present in this set: Communication (of policies) comm$(a \Rightarrow b, \phi)$ and data creation creates$(a, d)$. (Here $a, b$ are agents and $\phi$ is a ground policy formula, as introduced in the next subsection). Our system supports the addition of user-defined actions.

### 2.2 The Policy Language

Policies are used to express permissions that agents have, such as the permission to read a specific piece of data. Some requirements may guard a permission. These requirements can be *conditions*, as in 'Alice may read the data if she is a partner of Bob', or *obligations*, as in 'Alice may read the data if she pays Bob 10$'. Besides this, a policy may express or relate several different permissions. To provide maximum flexibility for writing policies, we now introduce the following policy language.

**Definition 1** *The set of* policies $\Phi$, *ranged over by $\phi$ and $\psi$, is defined by the following grammar:*

$$\begin{aligned}
\phi &::= p(s_1, ..., s_n) \\
&\quad \mid A \text{ owns } D \mid A \text{ says } \phi \text{ to } B \\
&\quad \mid \phi \wedge \phi \mid \forall x.\phi \mid \phi \rightarrow \phi \mid \xi \rightarrow \phi \\
s &::= A \mid D \\
\xi &::= !act \mid ?act
\end{aligned}$$

First, a policy formula can be a simple predicate $p(s_1, ..., s_n)$, where $s_i$'s can be either an agent, an agent variable, a data object or a data object variable. Second, we have the $A$ owns $D$ formula, which indicates that $A$ is the owner of data object $D$. As we define below, an data owner is allowed to create usage policies related to that data. $A$ says $\phi$ to $B$ expresses that agent $A$ is allowed to give policy $\phi$ to agent $B$. The 'says' policy contains a target agent to which the statement is said (different from e.g. [7, 1]). This allows us to provide a precise way of communicating policies to certain agents. However, the policy $A$ says $\phi$ to $B$ carries a different meaning for source agent $A$ than target agent $B$: While for agent $A$ it represents the permission to send $\phi$ to $B$, for $B$ it represents the possibility to use policy $\phi$ and delegate the responsibility to $A$.

The logic constructions *and*, *implication* and *universal quantification* have their usual meaning. We actually have two different instances of the implication. The first, $\phi' \rightarrow \phi$, has a policy $\phi'$ as a *condition*, stating that the agent first needs to establish this permission or fact before gaining the permission described in $\phi$. The second, $\xi \rightarrow \phi$, is used to express obligations. The requirement $\xi$ contains an action that the agent has to perform when the permission granted by $\phi$ is used. The annotations ! and ? indicate whether the agent needs to do this action every time it uses $\phi$ or only once. This will be discussed in Section 3.3. We write $\phi[D]$ to indicate that the set $D$ is the *data set of* $\phi$, i.e. all data objects and data variables occurring in $\phi$. For instance, we have read$(b, d)[\{d\}]$.

**Example 1** *The (atomic) policy that allows Bob to read the data $d$ is* read$(b, d)$.

1. *The policy that allows Bob to read every data object owned by Alice is* $\forall x.(a \text{ owns } x \rightarrow \text{read}(b, x))$.

2. *Let* $\text{age}\,21(x)$ *denote that agent* $x$ *is at least* 21 *years old, and* $\text{alc}(y)$ *denote that beverage* $y$ *is alcoholic. A policy allowing people over* 21 *to drink alcoholic beverages is* $\forall x, y.(\text{age}\,21(x) \wedge \text{alc}(y)) \rightarrow \text{drink}(x, y)$.

3. *If we require a payment of* 10$ *on the previous permission, the policy becomes* $\forall x.(!paid(x, 10\$) \rightarrow \forall y.(\text{age}\,21(x) \wedge \text{alc}(y)) \rightarrow \text{drink}(x, y))$.

## 2.3 Actions and permissions

To distinguish different instances of an action executed in the system, we label each instance using a unique identifier $id$, as in $\text{creates}_{id}(a, d)$. This formally gives a set $AC = \mathbb{N} \rightarrow ACT$ of 'executed actions' or 'action instantiations'. However, when possible, we simply talk about (labeled) actions in $AC$.

Three properties of actions that play a role in our policy system are described by the following functions:

- The *observability* function: $obs : AC \rightarrow P(\mathcal{G})$ describes which agents can observe which actions.

- The *proof obligation* function: $po : (AC \times \mathcal{G}) \rightarrow \Phi \cup \{\bot\}$ describes which policy an agent needs to justify the execution of an action. Here $\bot$ indicates that no policy is needed.

- The *conclusion derivation* function: $concl : (ACT \times \mathcal{G}) \rightarrow \Phi \cup \{\bot\}$, describes what policy can an agent deduce after observing an action. Here $\bot$ indicates that no policy can be deduced.

While the observability and proof obligation functions depend on executed actions (i.e. with identifiers), the conclusion derivation function is purely syntactical.

For our default actions $\text{creates}(a, d)$ and $\text{comm}(a \Rightarrow b, \phi)$ we have:

$$obs(\text{creates}(a, d)) = a \quad (1)$$
$$obs(\text{comm}(a \Rightarrow b, \phi)) = \{a, b\} \quad (2)$$
$$po(\text{creates}(a, d), a) = \bot \quad (3)$$
$$po(\text{comm}(a \Rightarrow b, \phi), c) = \bot \quad (a \neq c) \quad (4)$$
$$po(\text{comm}(a \Rightarrow b, \phi), a) = a \text{ says } \phi \text{ to } b \quad (5)$$
$$concl(\text{creates}(a, d), a) = a \text{ owns } d \quad (6)$$
$$concl(\text{comm}(a \Rightarrow b, \phi), b) = a \text{ says } \phi \text{ to } b \quad (7)$$
$$concl(\text{creates}(a, d), b) = \bot \quad (b \neq a) \quad (8)$$
$$concl(\text{comm}(a \Rightarrow b, \phi), c) = \bot \quad (c \neq b) \quad (9)$$

A creation action by $a$ is observed by $a$ (1), while a communication between $a$ and $b$ is observed by both $a$ and $b$ (2). In other settings, there may also be other agents that observe these actions, e.g. a router standing in between $a$ and $b$. Agents do not need a policy for creating data (3) or receiving a transmission (4). However, sending a transmission *does* require a permission (5). If agent Alice creates a piece of data she becomes the owner of this data (6); any other agent can not deduce the ownership (8). If an agent receives a communication then the agent can conclude the corresponding $says$ statement (7). However, any other agent can not deduce any conclusion (9).

**Remark 1** *Our communication* $\text{comm}(a \Rightarrow b, \phi)$ *models a point-to-point communication. We can easily model broadcasting, by introducing an action* $\text{bcast}(a, \phi)$, *and setting:*

$$obs(\text{bcast}(a, \phi)) = \mathcal{G}$$
$$po(\text{bcast}(a, \phi), x) = \begin{cases} \psi & (x = a) \\ \bot & \textit{otherwise} \end{cases}$$
$$concl(\text{bcast}(a, \phi), x) = \psi$$

Where $\psi = \forall y.a \text{ says } \phi \text{ to } y$.

Here, every agent can observe an action $\text{bcast}(a, \phi)$ and conclude that $a$ has broadcasted $\phi$ i.e. said $\phi$ to everybody. Only $a$ needs to justify this action.

## 2.4 The Proof System

In the previous section we introduced the actions that agents can execute and the permissions that agents need to justify these actions, in form of policies. This section describes how agents perform this justification, i.e. how agents can build policies from (simpler) ones. The possibilities for constructing policies are given in the form of a *derivation system* or *proof system* for our policy language.

Each rule includes, besides the *premises* and *conclusion*, an agent $a$, called the *context* of the proof, indicating which agent is doing the reasoning. Our derivation system DER contains the standard predicate logic rules for introduction and elimination of conjunction, implication and universal quantification, together with the following rules:

$$\text{SAY} \, \frac{b \text{ says } \phi \text{ to } a}{\phi} \, a$$

$$\text{REFINE} \, \frac{\phi \to \psi \quad a \text{ says } \phi \text{ to } b}{a \text{ says } \psi \text{ to } b} \, a$$

$$\text{OBS\_ACT} \, \frac{act \quad concl(act, a) \neq \bot}{concl(act, a)} \, a$$

$$\text{DER\_POL} \, \frac{a \text{ owns } d_1 \quad \ldots \quad a \text{ owns } d_n}{\phi[\{d_1, \ldots, d_n\}]} \, a$$

Rule (SAY) models delegation of responsibility. If agent $b$ says $\phi$ to $a$ then $a$ can assume $\phi$ to hold. (It is $b$'s responsibility to show that it had permission to give $\phi$ to $a$, see Section 3.3 on accountability.) Although agent $a$ may use $\phi$ without further requirement, it does not mean that the agent must always do this. If Bob wants to do a specific sensitive action, he may only want to use communications that he 'trusts' in building his policy. For example, Bob would only trust and thus use a policy 'fire Charlie' if it is provided by his boss. If it is provided to him by coworker Alice, Bob will not use the policy, even though the responsibility of this policy would rest with Alice. In this setting, the problem of establishing and managing trust is orthogonal to the problem of obtaining policies: One could introduce a trust management system to assign a 'level of trust in a proof', and require that different levels of trust are established for different actions (see Section 6).

In our logic, Alice can *refine* her own policies, e.g. by adding extra conditions and obligations using the standard propositional rules. In addition, rule (REFINE) enables Alice to refine the policies she provides to other agents: if Alice is allowed to send some policy $\phi$ then she can also send any refinement of $\phi$. This notion of refinement isn't easily captured in natural deduction. The current notation is too general; without further restrictions it seems that $a$ can derive $\phi \to \psi$ for any (syntactically unrelated) $\phi$, provided she can assume $\psi$. Thus, we need to restrict the ability to use assumptions in the subproof of $\phi \to \psi$. In the formalization of our model, a double context sequent calculus, we *can* express this (see Appendix A).

Rule (OBS\_ACT) links an action with its conclusion, given by the *concl* function. (OBS\_ACT) applies when there is some conclusion (i.e. $concl(act, a) \neq \bot$); e.g., from observing action $\text{comm}(a \Rightarrow b, \phi)$ $b$ derives $a$ says $\phi$ to $b$.

As we already mentioned, we design the logic in such a way that the owner of some data $d$ decides who is allowed to do which actions on $d$. In other words, an owner of some data $d$ is allowed to derive usage policies for $d$, targeted to any other agent. This is achieved by rule (DER\_POL), which allows the creation of any usage policy for data which the agent owns. Non-owners can refine existing policies (e.g., policies they received), but cannot create new policies from scratch.

A derivation with these rules made by an agent is a *proof*.

**Definition 2** *A proof $\mathcal{P}$ of $\phi$ from agent $a$ is a finite derivation tree such that: (1) each rule of $\mathcal{P}$ has $a$ as subject; (2) each rule of $\mathcal{P}$ belongs to DER, (3) the root of $\mathcal{P}$ is $\phi$, and (4) each initial assumption is either an action, an obligation or a basic predicate.*

*We call* conditions $cond(\mathcal{P})$ *of $\mathcal{P}$ the initial assumptions that are basic predicates, and* actions $act(\mathcal{P})$ *the initial assumptions which are observed unguarded actions (from rule (OBS\_ACT)). Finally, the multiset of initial assumptions that are guarded (by ? and !) actions are called the obligations $oblig(\mathcal{P})$ of $\mathcal{P}$.*

We now illustrate the usage of rules (REFINE) and (DER\_POL) in the following example.

$$
\begin{array}{c}
\text{REFINE} \dfrac{
\forall \text{I} \dfrac{
\rightarrow\text{I} \dfrac{
\rightarrow\text{I} \dfrac{[\,\mathsf{print}(b,d)\,]\ [\,\mathsf{rel}(d,x)\,]}{\mathsf{print}(b,d)}}{\mathsf{rel}(d,x) \rightarrow \mathsf{print}(b,d)}\ ^a
}{\dfrac{\forall x.\, \mathsf{rel}(d,x) \rightarrow \mathsf{print}(b,d)}{\mathsf{print}(b,d) \rightarrow \forall x.\, \mathsf{rel}(d,x) \rightarrow \mathsf{print}(b,d)}\ ^a}\quad
\text{OBS\_ACT} \dfrac{\mathsf{creates}(a,d)\quad concl(\mathsf{creates}(a,d),a)) = a\ \mathsf{owns}\ d}{\text{DER\_POL}\dfrac{a\ \mathsf{owns}\ d}{a\ \mathsf{says}\ \mathsf{print}(b,d)\ \mathsf{to}\ b}\ ^a}\ ^a
}{a\ \mathsf{says}\ \forall x.\, \mathsf{rel}(d,x) \rightarrow \mathsf{print}(b,d)\ \mathsf{to}\ b}\ ^a
$$

**Table 1. Proof for Example 2.**

**Example 2 (Policy Refinement)** *Suppose we have a predicate* $\mathsf{rel}(d, \bar{d})$, *expressing whether two data objects are related: For instance, $d$ can be a review of a new product and $\bar{d}$ the press release announcing this product. Alice creates $d$ and wants to give a policy $\forall x.\, \mathsf{rel}(d,x) \rightarrow \mathsf{print}(b,d)$ to Bob giving Bob permission to print the document as soon as a related object exists: Alice can build the policy allowing her to give this policy to Bob as shown in Table 1.*

## 3 The Model

We now introduce a model for our system, combining the different components of the previous sections. In our system, agents can execute and log actions. In addition to agents, an authority is also present. This authority may audit agents requiring justification for (some of) the agents actions.

### 3.1 Logging actions

Whenever an agent executes an action, it can also choose to *log* this action. Logged actions constitute evidences that can be used to demonstrate that an agent was allowed to perform a particular action. They are used during accountability auditing, in Section 3.3.

**Definition 3** *A* logged action *is a triple $lac = \langle act, conds, obligs \rangle$ consisting of an action $act \in AC$, a set of atomic predicates $conds$ (the 'conditions'), and a set of labelled annotated actions $obligs \subset \{!, ?\}AC$ (the 'obligations'). The set of logged actions is denoted as $LAC$.*

When logging an action, an agent can include supporting conditions which the environment certifies to be valid at the moment of execution of the action. This is recorded in the set of predicates $conds$. We do not model the environment explicitly but instead assume that the agent obtains a secure "package" of signed facts from the environment, represented in $conds$. As an example, one can think of the driver's license of Alice being checked to certify that she is over 21.

An agent can also include obligations in $obligs$ in a logged action, which refers to other actions the agent did or promises to do. We abstract away from the details of expressing promises, and instead assume we have a way to check if actions have *expired*. For example, the agent may promise to pay within a day. Then a payment action needs to be done (and logged) within a day of logging this obligation. (Also see Section 3.3.)

**Example 3** *We continue with Example 1.3. Suppose that we introduce an action $\mathsf{drunk}(x,y)$ and a corresponding atomic predicate $\mathsf{drink}(x,y)$, with $concl(\mathsf{drunk}(x,y),x) = \bot$ and $po(\mathsf{drunk}(x,y),x) = \mathsf{drink}(x,y)$. We also introduce an action $\mathsf{paid}(x,y)$, with corresponding atomic predicate $\mathsf{pay}(x,y)$, with $concl(\mathsf{paid}(x,y),x) = po(\mathsf{paid}(x,y),x) = \bot$.*

*A logged action $lac_{\mathsf{pay}}$ for payment is done first by $a$:*

$$lac_{\mathsf{pay}} = \langle \mathsf{paid}_0(a, 10\$), \emptyset, \emptyset \rangle$$

*Then, another logged action $lac_{\mathsf{drunk}}$ for the action $\mathsf{drunk}_1(a, beer)$ is recorded:*

$$
\begin{aligned}
lac_{\mathsf{drunk}} =\ & \langle \mathsf{drunk}_1(a, beer),\\
& \{\mathsf{age21}(a), \mathsf{alc}(beer)\},\\
& \{!\mathsf{paid}_0(a, 10\$)\} \rangle
\end{aligned}
$$

The *log of an agent* $a$ is a finite sequence of logged actions. Note that it does not need to be $a$ who performed the actions, but of course $a$ has to observe an action to be able to log it. We say that agent $a$ logs action $act$ when $\langle act, conds, obligs \rangle$ is appended to the log of $a$, where $conds$ is some set of conditions and $obligs$ is some set of obligations.

We assume the following consistency properties of logging:

- An agent logs any action at most once, thus within an agent's log the logged actions are uniquely identified by the label (id) of the action.

- An agent can include the same obligation $!act_{id}$ at most once within the obligations of logged actions in its log. (an $?act_{id}$ action, in contrast, may occur multiple time).

- An agent cannot log an expired action.

Notice that consistency of the log does not have to be checked at time of logging, it is sufficient to check it at time of auditing.

## 3.2 The system model and state

We are ready to introduce our system model.

**Definition 4** *A system is a 6-tuple:*

$$\langle \mathcal{G}, \Phi, ACT, obs, concl, po \rangle$$

*where $\mathcal{G}$ is a set of agents, $\Phi$ is the policy language, $ACT$ is a set of actions, and $obs$, $concl$ and $po$ are, respectively, the observability, conclusion and proof obligation functions.*

A *state* $\mathcal{S}$ is the collection of logs of the different agents, i.e. a mapping from agents to logs. An agent who observes an action may choose to log this action. Thus by executing action $act$ the system can make a *transition* from a state $\mathcal{S}$ to state $\mathcal{S}'$, denoted $\mathcal{S} \stackrel{act}{\rightarrow} \mathcal{S}'$ where $\mathcal{S}'$ equals $\mathcal{S}$ except that the action $act$ may have have been logged by any agent $a$ that can observe the action, $a \in obs(act)$. An *execution* of the system consists of a sequence of transitions $\mathcal{S}_0 \stackrel{act_1}{\rightarrow} \mathcal{S}_1 \stackrel{act_2}{\rightarrow} ... \stackrel{act_n}{\rightarrow} \mathcal{S}_n$, starting with some initial state $\mathcal{S}_0$. The *execution trace* for this execution is $act_1 \, act_2 ... \, act_n$. Actions logged by an agent can be also seen as a trace of actions, by projecting only the actions of each logged action. We denote that trace as $\mathcal{S}(a)$. Let $\preceq$ denote the subtrace relation ($tr_1 \preceq tr_2$ if each action of $tr_1$ is included in $tr_2$, and each time an action $act_1$ appears before $act_2$ in $tr_1$, the $act_1$ also appears before $act_2$ in $tr_2$). We have $\mathcal{S}(a) \preceq tr$.

**Auditing Authority** Agents may be audited by some *authority*, at some state $\mathcal{S}$. Intuitively, when some agent is about to be audited, an auditing authority is formed. This authority will audit the agent to find whether she is accountable for her actions. Let $tr$ be the sequence of actions executed from some initial state to $\mathcal{S}$, The *evidence* trace, denoted $\mathcal{E}$, contains all the actions that might be audited. Initially, $\mathcal{E}$ embeds $\mathcal{S}(a)$. However, $\mathcal{E}$ may also contain actions not in $\mathcal{S}(a)$: They may be provided, for example, by some observing agents. However, we assume that given $\mathcal{S}(a)$ and other observed actions $S$, the authority can order properly the actions of $\mathcal{S}(a)$ and $S$ into $\mathcal{E}$, s.t. $\mathcal{E} \preceq tr$. Thus, in general $\mathcal{E}$ is a trace satisfying $\mathcal{S}(a) \preceq \mathcal{E} \preceq tr$.

## 3.3 Accountability

We now introduce notions of agent accountability, determined by some authority in possession of evidences. These definitions allow an authority to audit agents, to establish whether the agent was allowed to do the actions he did. In previous work [6], we defined several notions of agent and data accountability, but without checking for obligations nor conditions. We did not have logs of agents either. We now define accountability for logged actions, which we then extend to agent logs.

We first introduce *justification* proofs for logged actions. Intuitively, a justification proof is a proof of the policy required for the action (as given by function $po$), using only conditions and obligations that have been logged.

**Definition 5** *A proof $\mathcal{P}$ of $\phi$ from $a$ is a justification (proof) of logged action $\langle act, conds, obligs \rangle$ if:*

- $po(act, a) = \phi$

- *The obligation in the proof are included in obligs; 'oblig($\mathcal{P}$) $\subset$ obligs'. (Here multiple ?act in oblig($\mathcal{P}$) may be assigned to the same ?act$_{id}$ but each occurrence of !act must have its own !act$_{id}$ in obligs.)* [1]

- *Each condition in the proof is in conds; cond($\mathcal{P}$) $\subseteq$ conds.*

*The set of all justifications is denoted by $\mathcal{J}$.*

In general, there may be different justifications for an action. The justifications provided by the agent are modeled by a function $\text{Pr} : \mathcal{G} \times LAC \to \mathcal{J} \cup \{\bot\}$. Here $\text{Pr}(a, \langle act, conds, obligs \rangle)$ is either a valid justification of $\langle act, conds, obligs \rangle$, or it is $\bot$, indicating that the agent did not provide a justification.

**Definition 6 (Logged Action Accountability)** *Agent $a$ correctly accounts for logged action $logact = \langle act, conds, obligs \rangle$ (in state $\mathcal{S}$), denoted $LAA(a, logact)$, if:*

- *if $po(act, a) \neq \bot$ then $\text{Pr}(a, logact) \neq \bot$, i.e. if needed a justification is provided*

- *if $o \in obligs$ has expired then $o \in \mathcal{S}(a)$, i.e. each obligation that has expired has been (executed and) logged.*

- *For each $act \in act(\text{Pr}(a, logact))$, $a$ provides an id s.t. $act_{id}$ occurs in $tr$ and $a \in obs(act_{id})$*[2]

This definition introduces accountability for a single (logged) action. We now define accountability for any action and for all audited actions.

**Definition 7 (Action Accountability)** *We say that agent $a$ correctly accounts for (labeled) action $act$, denoted $AA(a, act)$, if*

- *$a$ has logged $act$ as $logact$ and $LAA(a, logact)$ or*

- *$a$ has not logged $act$ and $LAA(a, \langle act, \emptyset, \emptyset \rangle)$*

*We say agent $a$ passes audit $\mathcal{E}$, written $ACC(a, \mathcal{E})$, if either $\mathcal{E}$ is the empty trace, or $\mathcal{E} = \mathcal{E}'.act$ with:*

- *$AA(a, act)$*

- *$ACC(a, \mathcal{E}'')$, with $\mathcal{E}''$ the correct ordered merge of $\mathcal{E}'$ and $newacts$, all the* new *actions (i.e. not already in $\mathcal{E}'$) given by the proofs in $AA(a, act)$.*

The second case for action accountability explains why it can be in the interest of an agent to log its actions. As conditions may have changed, the agent can only rely on conditions if they have been logged at the time the action was executed. For example, if some action had a condition 'only execute between 4 p.m. and 4:30 p.m. on 12/10/2004', then that condition would only hold temporarily; If an agent executed the action and did not log it, during a later audit the agent could not provide a valid proof. The same holds for obligations: Only obligations logged with the action can be used in a proof of the action.

**Claim 1** *Let $a$ be an agent and $\mathcal{E}$ an evidence trace. Then, checking $ACC(a, \mathcal{E})$ terminates.*

*Proof.* Let $|tr| = n$, for some $n \geq 0$. Suppose $|\mathcal{E}| = m \leq n$. We show that each execution of $ACC(a, \mathcal{E})$ decreases $l$, with $l = n$ initially. After every execution of $ACC(a, \mathcal{E})$ as in Definition 7, at most $l - m$ evidences (the $newacts$) are added to $\mathcal{E}''$. Thus, $|\mathcal{E}''| = |\mathcal{E}'| + (l - m) = m - 1 + (l - m) = l - 1$. Hence, $ACC(a, \mathcal{E})$ terminates.

**Honest Strategy** A strategy for an honest agent $a$ to always be accountable is as follows. Before executing some action $act$, $a$ checks whether $po(act, a)$ is derivable. If any obligation needs to be fulfilled, then the agent performs and then logs it. If any condition or obligation needs to be fulfilled, then the action $act$ is also logged. Then, it follows from the definitions that:

**Remark 2 (Accountability of honest agents)** *If agent $a$ follows the* honest strategy, *then for any system execution and any auditing authority with evidence set $\mathcal{E}$, we have that $ACC(a, \mathcal{E})$ holds.*

The proof follows immediately from Definitions 6 and 7.

---
[1] interpreting as sets for ?act and as multisets for !act.
[2] We assume the authority can verify this.

**Recursive auditing** We have, up to now, defined accountability of one particular agent in isolation. However, we may be interested in cross-verifying the actions of agents. We sketch an algorithm for recursive auditing of agents, which can be used by a potential auditing authority. The algorithm inputs $S_0$, a set of *suspected* agents, and $\mathcal{E}_0$, an initial evidence trace. Given $ACC(a, \mathcal{E})$, we write $E(ACC(a, \mathcal{E}))$ to denote the set of new actions appearing in the given proofs (the $newacts$ in Definition 7), and $A(ACC(a, \mathcal{E}))$ the corresponding set of agents appearing in these actions for which the proof obligation is not bottom, i.e. $po(\cdot, \cdot) \neq \perp$.

**Algorithm 1 (Recursive Auditing)** *Inputs: $S_0$ and $\mathcal{E}_0$. Outputs:* **true** *if audited agents are accountable,* **false** *otherwise.*

1. $S := S_0$;
2. $\mathcal{E} := \mathcal{E}_0$;
3. **while** $S \neq \emptyset$ **do**
4.   **let** $a \in S$;
5.   **if** $ACC(a, \mathcal{E})$ **then**
6.     $S := (S \setminus \{a\}) \cup A(ACC(a, \mathcal{E}))$
7.     $\mathcal{E} := \mathcal{E} \cup E(ACC(a, \mathcal{E}))$
8.   **else**
9.     return **false**
10. **end**
11. return **true**

**Claim 2** *Algorithm 1 terminates.*

*Proof.* Similar to the proof of Claim 1.

**Claim 3** *In line 4 of of Algorithm 1, the order in which the agents are chosen does not matter.*

*Proof (sketch).* Follows from the fact that proofs are fixed on beforehand, as given by function $\text{Pr} : \mathcal{G} \times LAC \to \mathcal{J} \cup \{\perp\}$, and do not depend on knowing whether other agents are being audited or not. (Intuitively, this models the fact that agents can not change their proofs on the fly, depending on whether other agents are being audited.)

## 4 Formalization

We have implemented the proof system and checking of the audit logic in Twelf [13], which is an implementation of the Edinburgh Logical Framework [12]. Research in proof-carrying code [10] has shown that Logical Framework (LF) provides a suitable notation for proofs to be sent and checked by a recipient. In type theories proof checking reduces to type checking and the LF proof checker is as simple as a programming language type checker (see also Section 5).

Here we only mention the most important features of the implementation. In the appendices we have included both a more abstract representation (common for a sequent calculus), as well as the complete Twelf-file.

**Audit Logic Implementation** We first declare the types of the object logic, the atomic predicates and the actions. Then we present the proof rules and we finish with an example of a complete proof.

**Types** In Twelf, a metalogic type is of type `type`. For object logic types, we use the type `tp`. The meta-logic function `->` goes from `type` to `type` or `kind` (`type` has type `kind`). Agents, data, actions and policies are declared as `tp`s. Finally, when declaring the rules for the object logic the `tm` type constructor is used, which casts arguments from `tp` to `type`. Summarizing:

```
tp: type.         agent: tp.     policy: tp.
tm: tp -> type.   data: tp.      action: tp.
```

**Policies** Recall that policies are formed with `says` and `owns` and for example scenario-specific atomic predicates like `print`:

```
says: tm agent -> tm policy -> tm agent -> tm policy.
owns: tm agent -> tm data -> tm policy.
print: tm agent -> tm data -> tm policy.
```

For instance, (print a d) is a policy stating that agent a has permission to print document d. Policies can also be formed using the usual propositional connectives and universal quantification[3](except for negation and disjunction); to model the different instances of implication in the policy language (see definition 1), we declare separate instances for use-once-obligations and use-many-obligations:

```
imp: tm policy -> tm policy -> tm policy.
forall: (tm T -> tm policy) -> tm policy.
and: tm policy -> tm policy -> tm policy.
?imp: tm action -> tm policy -> tm policy.
!imp: tm action -> tm policy -> tm policy.
```

Below we assume that the connectives imp, and, ?imp, !imp are declared as infix-operators.

**Actions**   The actions creates and comm have the following types:

```
creates: tm agent -> tm data -> tm action.
comm: tm agent -> tm agent -> tm policy -> tm action.
```

The *conclusion derivation* function describes the policies the agent can deduce from some action.

```
concl: tm action -> tm agent -> tm policy -> type.
concl_comm: concl (comm A B Phi) B (says A Phi B).
concl_creates: concl (creates A D) A (owns A D).
```

In Twelf, symbols with a leading capital are variables. Their type can often be left unspecified, as Twelf expands them to the most general type.

**Proof derivation**   To model local proofs (with respect to agents), we use sequent calculus formulas of the form $\Gamma; \Delta \vdash_A \Phi$, indicating that agent $A$ can deduce the policy $\Phi$ from premises $\Gamma$ and $\Delta$. Here $\Gamma$ is an unrestricted context and $\Delta$ a linear context. More precisely, $\Gamma$ is a sequent of policies, actions and ?obligations, while $\Delta$ contains only !obligations. We formalize this with

```
entail: tm agent -> list nonlin -> list lin -> tm policy -> type.
```

where list T is the type for lists of type T. The type nonlin is like a disjoint union (policy, action, action) by the following definitions:

```
nonlin: tp.
lin: tp.
act_c: tm action -> tm nonlin.
?act_c: tm action -> tm nonlin.
pol_c: tm policy -> tm nonlin.
!act_c: tm policy -> tm lin.
```

**Rules**   The rules SAY and REFINE are defined as follows:

```
say: entail B (cons (pol_c Phi) Gamma) Delta Psi ->
     entail B (cons (pol_c (says A Phi B)) Gamma) Delta Psi.
```

The OBS_ACT rule works as follows. If an agent $A$ can conclude the policy $\Phi$ by observing action *act*, then she can deduce $\Phi$ from any set $\Gamma$ containing *act*. In formula $\Gamma, act; \Delta \vdash_A \Phi$, in Twelf:

```
obs_act:  concl Act A Phi ->
          entail A (cons (act_c Act) Gamma) Delta Phi.
```

The REFINE rule implements the possibility of *refining* a policy, $\phi$ to $\psi$ and saying $\psi$, if one has permission to say $\psi$. Refinement in this sense is expressed using empty sequents:

```
refine:  entail A (cons (pol_c Phi) nil) nil Psi ->
         entail A (cons (pol_c (says A Phi B)) Gamma) Delta (says A Psi B).
```

---
[3]Quantification over policies is not allowed, so we only instantiate the ∀-left and -right rules for the types agent and data.

The meaning of the DER_POL rule is that any formula can be derived, as long as, for all data affected by it, ownership is proven. To this end we define a relation on a non-linear sequent $\Gamma$ and a policy $\phi$, op$[\Gamma; \phi]$, that only holds iff all data affected by $\phi$, occur in some owns-predicate in $\Gamma$.

```
op: list nonlin -> tm policy -> type.
op_imp: (op Gamma Phi) -> (op Gamma (Psi imp Phi)).
op_forall: ({X: tm data} op Gamma (Phi X)) -> (op Gamma (forall Phi)).
op_says: (op Gamma Phi) -> (op Gamma (says B Phi C)).
op_owns: (op (cons (pol_c (owns A D)) Gamma) (owns B D)).
op_print: (op (cons (pol_c (owns A D)) Gamma) (print B D)).
```

Now the DER_POL rule is defined as follows:

```
der_pol: (op Gamma Phi) ->
         entail A Gamma Delta Phi.
```

Please note that the structural rules and the logical rules for conjunction, implication and universal quantification over agents and data are omitted here, but they can be found in the appendices.

**Example 4 (Formalization of Example 2)** *Recall that Alice creates a document* d *(a review of a product) and communicates a policy, to Bob, that gives Bob permission to print* d *'as soon as a related object exists'. Below Alice justifies that she can do so. First we declare the scenario-specific objects and predicates.*

```
a: tm agent. b: tm agent. d: tm data.
rel: tm data -> tm data -> tm policy.
phi = forall [x:tm data]((rel d x) imp (print b d)).

ex2: entail a (cons (act_c (creates a d)) nil) nil (says a phi b)=
     obs_act concl_creates
        (cut
           (der_pol (op_says op_print))
           (refine
              (forall_r_data [x] (imp_r (w_l init))))
           append_nil2 append_nil).
```

This theorem has been checked by Twelf. The proof goes along the same lines as previously in Table 1.

The lhs can be expressed as $\gamma; \delta \vdash_a (a \text{ says } \phi \text{ to } b)$, where $\phi$ is $\forall x. \text{rel}(d, x) \to \text{print}(b, d)$ and $\gamma$ only contains the creates action, while $\delta$ is empty. In the proof above, `imp_r` is the $imp-right$ rule, `forall_r_data` is the $\forall-right$ rule for quantification over data, `w_l` and `init` are *weakening-left* and the *initial sequent axiom* for the non-linear sequent.

## 5 Related Work

There is a wide body of literature on logics in Access Control (see the survey by Abadi [1]). Here, we mention some of the proposals. Binder [7] is a logic-based security language based on Datalog Binder includes a special predicate, *says*, used to quote other agents. Binder's *says* differs in two aspects from our construct: First, ours includes a target agent (see Section 2.4); Second, when importing (i.e. communicating a policy in our setting) a clause in Binder, care must be taken to avoid nested *says*, since it may introduce difficulties in their setting. More related to our auditing by means of proofs, Appel and Felten [2] propose the Proof-Carrying Authentication framework (PCA), also implemented in Twelf (see Section 4). Differently from our work, PCA's language is based on a higher order logic that allows quantification over predicates. Also, their system is implemented as an access control system for web servers, while in our case we focus on a-posteriori auditing.

BLF [19] is an implementation of a Proof-Carrying-Code framework that uses both Binder and Twelf, which however focuses on checking semantic code properties of programs.

Sandhu and Samarati [16] give an account of access control models and their applications. Bertino et al. [4] propose a framework for reasoning on access control models, in which authorization rules treat the core components Subjects, Objects and Privileges. Sandhu and Park [11] take a different approach with their UCON-model, in which the decision is modelled as a reference monitor that checks the 3 components: ACL, Conditions and Obligations. This reflects much the separation also made by us. Obligations and conditions are also prominent in directives on privacy and terms of use in DRM. The concept of *purpose* of an action is not used by us, but is used in the privacy languages P3P and E-P3P [3]. Unlike our policy language, E-P3P allows the use of negation, which requires special care to avoid problems in a distributed setting.

# 6  Conclusions and Future Work

We have presented a flexible usage policy framework which enables expressing and reasoning about policies and user accountability. Enforcement of policies is difficult (if not impossible) in the highly distributed setting we are considering. Instead, we propose an auditing system with best-effort checking by an authority depending on the power of the authority to observe actions. A notion of agent accountability is introduced to express the proof obligation of an agent being audited.

Our obligations cover pre- and post-obligations ([15]) but not yet ongoing obligations. The setup does, with an adaption of the definitions of accountability, seem to provide the means to include this type of obligations. Obligations are 'use once', e.g. $!pay(\$10)$ or 'use as often as wanted' $?pay(\$10)$.

Our proof system has been implemented using the proof checker Twelf. The agents develop proofs using this implementation. Likewise, the implementation allows an authority to check the agents' proofs.

In our system, we include a powerful rule which allows delegating any policy to any other agent. Agent Alice may only want to use a policy from Bob if she (i) *knows* Bob, (ii) *authenticates* Bob, and (iii) *trusts* Bob. All these issues are (intentionally) abstracted away in our approach, as they seem to be orthogonal to our aims. For example, in (iii), the required level of trust may depend on the policy provided by Bob or on the way Alice is going to use the policy. There, a distributed trust management system (e.g. [9]) could be employed to obtain the required level of trust.

In the work of Samarati et.al. [14], a discussion about decentralized administration is presented. Specially, the revocation of authorizations is addressed. This is acomplex problem, which occurs as a consequence of the delegation of privileges. One could model revocation of policies by adding a special flag plus a corresponding check in a policy. However, checking whether a flag is set in another agent's environment is not realistic in our highly distributed setting. Further research is needed to find a both practical and realistic way to include rights revocation.

Our implementation only covers proof checking. Revisiting Figure 1, we find that arrows covering policy communications and logging are not yet implemented. Our framework requires several properties for each of the different modules (e.g. secure logging and non-repudiable communications). Certainly, these properties need cryptography to be realized securely. We regard as future work the rigorous cryptographic definition of these properties, along with the accompanying cryptographic constructions.

**Acknowledgements**   The research presented in this paper is conducted within the PAW project (funded by Senter-IOP and TNO), the Inspired Project and the NWO Account Project.

## A  The Sequent calculus for the Audit Logic

In Twelf we implemented a double context sequent calculus. To explain the Twelf-code more clearly, we start by giving the (for sequent calculi) common representation of our system. In the formalization of our model we have in general ⟨context⟩; ⟨linear context⟩ $\vdash_A$ policy , **where $A$ is the agent doing the reasoning, but below we can write shortly** $\Gamma; \Delta \vdash \phi$ (because rules regard only one agent-context). The contexts are sequents of linear and nonlinear propositions that serve as premises for the conclusion (right of the $\vdash$ sign). Now, in the Twelf code we declare 6 types of objects: agent, data, action, policy, nonlin, lin and a type list. In the sequel Greek and Roman letters denote variables that range over them. The type of the variables is often left implicit, so to distinguish the different types we use; A, B, C to denote agents, $\Gamma$, $\Gamma'$ for lists of type nonlin, $\Delta$, $\Delta'$ for lists of type lin, $\gamma_0$ and $\delta_0$ for the empty lists of type lin and nonlin. Policies are denoted with $\phi$, $\phi'$, $\psi$ and $\xi$, $\xi'$ denote actions. To express append and construct operators for lists we just write a comma. And finally $\rightarrow, \wedge, \forall$ denote the policy connectives for implication, conjunction and universal quantification. We use `?imp` and `!imp` for the implication operators on obligations. Finally, in the non-linear context, we mix policy formulas (introduced by the standard logical rules), with actions (introduced by obs_act) and use-many-obligations (introduced with `?imp` ). In Twelf each of these is mapped to the type nonlin. This context is thus like a list of the disjoint union of policy, action and action.

Below we only mention the main inference rules. For the type declarations and the definitions of lists and so on we refer to the Twelf code in the appendix.

First our pivotal rules. Here we mention the $A$ in $\vdash_A$ because the agent appears explicitly in the formulas.

$$\text{SAY}\frac{\Gamma, \phi; \Delta \vdash_A \psi}{\Gamma, \mathsf{says}(\mathtt{B}, \phi, \mathtt{A}); \Delta \vdash_A \psi}$$

$$\text{OBS\_ACT}\frac{\Gamma, concl(\xi, \mathtt{A}); \Delta \vdash \psi}{\Gamma, \xi; \Delta \vdash_A \psi}$$

$$\text{REFINE}\frac{\gamma_0, \phi; \delta_0 \vdash_A \psi}{\Gamma, \mathsf{says}(\mathtt{A}, \phi, \mathtt{B}); \Delta \vdash_A \mathsf{says}(\mathtt{A}, \psi, \mathtt{B})}$$

$$\text{DER\_POL}\frac{\mathsf{op}[\Gamma; \psi]}{\Gamma; \Delta \vdash_A \psi}$$

Note the apparent difference with for instance the SAY-rule in natural deduction style in Section 2.4. It seems to have been turned upside-down; this is done to both keep the sub-formula-property, and preserve the redundancy of the cut-rule[4].

The OBS\_ACT-rule uses the conclusion derivation function, which is defined as:

$$\mathtt{concl\_creates}: concl(creates(A, D), A) = \mathsf{owns}(A, D).$$
$$\mathtt{concl\_comm}: concl(comm(A, B, \phi), B) = \mathsf{says}(A, \phi, B).$$

A few words on the REFINE-rule: The intention of the empty sequents is to prevent that an agent uses a fact $\psi$ to derive $\phi \to \psi$ for an otherwise unrelated $\phi$. In formulas: $\Gamma, \psi \vdash (\phi \to \psi)$ holds for any $\Gamma$ and $\phi$. Then with just the permission to *say* $\phi$, the agent can *say* $\psi$, which is of course not the meaning of *refining* the policy $\phi$. This subtlety is not easily expressed in natural deduction rules, but in the sequent calculus this is done straightforwardly with the empty sequent $\gamma_0$. With this restriction in place only syntactically related policies can be used, for example for $\psi = (\xi \to \phi)$ or $\psi = \forall \phi$. While, if there exists such a relation for two (syntactically unrelated) predicates, then it should be expressed by adding a rule for it: $\gamma_0, \mathsf{write}(A, D) \vdash \mathsf{read}(A, D)$.

Finally, the DER\_POL-rule expresses that if a policy sets permissions on data $D_1, ..., D_n$ and each of the $D_1, ..., D_n$ is owned by $A$ then $A$ can derive that policy. The notion of *dataset* of a policy in the article, is thus formalized a bit stronger here with the notion of *active dataset*, being the data that is actually affected by the policy. To this end we defined a relation on $\Gamma$ and $\phi$: $\mathsf{op}[\Gamma; \phi]$ that holds iff all data affected by $\phi$, occur in some owns-predicate in $\Gamma$. The following rules make up its definition:

$$\mathtt{op\_imp}\frac{\mathsf{op}[\Gamma; \phi]}{\mathsf{op}[\Gamma; (\psi \to \phi)]}$$

$$\mathtt{op\_forall}\frac{\mathsf{op}[\Gamma; \phi(\mathtt{X})]}{\mathsf{op}[\Gamma; \forall \phi]} \text{ (X a fresh constant)}$$

$$\mathtt{op\_says}\frac{\mathsf{op}[\Gamma; \phi]}{\mathsf{op}[\Gamma; \mathsf{says}(\mathtt{B}, \phi, \mathtt{C})]}$$

$$\mathtt{op\_owns}\frac{}{\mathsf{op}[\Gamma, \mathsf{owns}(\mathtt{A}, \mathtt{D}); \mathsf{owns}(\mathtt{B}, \mathtt{D})]}$$

$$\mathtt{op\_print}\frac{}{\mathsf{op}[\Gamma, \mathsf{owns}(\mathtt{A}, \mathtt{D}); \mathsf{print}(\mathtt{B}, \mathtt{D})]}$$

---
[4]A full proof of cut-admissibility and other meta-theorems are future work

Where the last one is to show a scenario-specific predicate. Now we present the standard rules for the sequent calculus:

$$\text{init} \frac{}{\Gamma, \phi; \Delta \vdash \phi}$$

$$\text{cut} \frac{\Gamma; \Delta \vdash \phi \quad \Gamma', \phi; \Delta' \vdash \psi}{\Gamma, \Gamma'; \Delta, \Delta' \vdash_A \psi}$$

$$\text{and\_l1} \frac{\Gamma, \phi; \Delta \vdash \phi'}{\Gamma, (\phi \wedge \psi); \Delta \vdash \phi'}$$

$$\text{and\_l2} \frac{\Gamma, \psi; \Delta \vdash \phi'}{\Gamma, (\phi \wedge \psi); \Delta \vdash \phi'}$$

$$\text{and\_r} \frac{\Gamma; \Delta \vdash \phi \quad \Gamma'; \Delta' \vdash \psi}{\Gamma, \Gamma'; \Delta, \Delta' \vdash (\phi \wedge \psi)}$$

$$\text{imp\_l} \frac{\Gamma; \Delta \vdash \phi \quad \Gamma', \phi'; \Delta' \vdash \psi}{\Gamma, \Gamma', (\phi \to \phi'); \Delta, \Delta' \vdash \psi}$$

$$\text{imp\_r} \frac{\Gamma, \phi; \Delta \vdash \psi}{\Gamma; \Delta \vdash (\phi \to \psi)}$$

We need similar left and right rules for the !imp and ?imp operators. Recall that use-many obligations go into the unrestricted context, while use-once obligations go into the linear context. Here are the two pairs:

$$\text{!imp\_l} \frac{\Gamma; \Delta \vdash (\xi \ !\text{imp} \ \phi)}{\Gamma; \Delta, \xi \vdash \phi}$$

$$\text{!imp\_r} \frac{\Gamma; \Delta \vdash \phi}{\Gamma; \Delta \vdash (\xi \ !\text{imp} \ \phi)}$$

$$\text{?imp\_l} \frac{\Gamma; \Delta \vdash (\xi \ ?\text{imp} \ \phi)}{\Gamma, \xi; \Delta \vdash \phi}$$

$$\text{?imp\_r} \frac{\Gamma; \Delta \vdash \phi}{\Gamma; \Delta \vdash (\xi \ ?\text{imp} \ \phi)}$$

In Twelf we coded twice the following ∀-left and ∀-right rules; one for agents, one for data. But not for actions or policies.

$$\text{forall\_l} \frac{\Gamma, \phi(X); \Delta \vdash \psi}{\Gamma, \forall \phi; \Delta \vdash \psi}$$

$$\text{forall\_r} \frac{\Gamma; \Delta \vdash \phi(X)}{\Gamma; \Delta \vdash \forall \phi} \ (X \text{ a fresh constant})$$

Finally, we have the structural rules for the unrestricted and the linear contexts. The linear context doesn't allow for contraction.

$$\text{w\_l} \frac{\Gamma; \Delta \vdash \phi}{\Gamma, \psi; \Delta \vdash \phi}$$

$$\text{w\_l\_act} \frac{\Gamma; \Delta \vdash \phi}{\Gamma; \Delta, \xi \vdash \phi}$$

$$\text{contr\_l} \frac{(\Gamma, \phi), \phi; \Delta \vdash \psi}{\Gamma, \phi; \Delta \vdash \psi}$$

$$\text{perm\_l} \frac{\Gamma, \phi, \phi', \Gamma'; \Delta \vdash \psi}{\Gamma, \phi', \phi, \Gamma'; \Delta \vdash \psi}$$

$$\text{perm\_act} \frac{\Gamma; \Delta, \xi, \xi', \Delta' \vdash \phi}{\Gamma; \Delta, \xi', \xi, \Delta' \vdash \phi}$$

Which wraps up the list of rules. We now show some examples of proofs in this notation.

## A.1 Example: Creating and consuming a policy

Let's draw the proof-tree of example 2a in this notation. We use abbreviations in this example:

$$\phi = \forall \psi \qquad \psi(x) = (\psi_1(x) \to \psi_2)$$
$$\psi_1(x) = \text{rel}(d, x) \qquad \psi_2 = \text{print}(b, d)$$
$$\gamma = \gamma_0, \text{owns}(a, d) \qquad \gamma' = \gamma_0, \psi_1(d')$$

In words: Agent $a$ proves that she can say policy $\phi$ to agent $b$.

$$\text{OBS\_ACT} \frac{\text{cut} \dfrac{\text{DER\_POL} \dfrac{\text{op\_says} \dfrac{\text{op\_print} \dfrac{}{\text{op}[\gamma; \psi_2]}}{\text{op}[\gamma; \text{says}(a, \psi_2, b)]}}{\gamma; \delta_0 \vdash \text{says}(a, \psi_2, b)} \quad \text{REFINE} \dfrac{\gamma_0, \psi_2; \delta_0 \vdash \phi}{\gamma_0, \text{says}(a, \psi_2, b); \delta_0 \vdash \text{says}(a, \phi, b)}}{\gamma; \delta_0 \vdash \text{says}(a, \phi, b)}}{\gamma_0, creates(a, d); \delta_0 \vdash \text{says}(a, \phi, b)}$$

Please note that the use of both the REFINE and the DER_POL-rule, and thus the cut-rule, is only done to stay close to the natural deduction style proof given earlier in the article (see Table 1). In fact, by using only the DER_POL-rule one could derive the same permission. The premise of the REFINE-rule is easily proven:

$$\text{forall\_r} \dfrac{\text{imp\_r} \dfrac{\text{w\_l} \dfrac{\text{init} \dfrac{}{\gamma_0, \psi_2; \delta_0 \vdash \psi_2}}{\gamma_0, \psi_2, \psi_1(y); \delta_0 \vdash \psi_2}}{\gamma_0, \psi_2; \delta_0 \vdash (\psi_1(y) \to \psi_2)}}{\gamma_0, \psi_2; \delta_0 \vdash \forall \psi}$$

Now agent $b$, who receives the mentioned policy, proves that he has the permission to print - using the evidence of the communication by $a$ and the assertion on $d'$ (inside $\gamma'$) which is asserted by $b$'s environment.

$$\text{OBS\_ACT} \dfrac{\text{SAY} \dfrac{\text{forall\_l} \dfrac{\text{imp\_l} \dfrac{\text{init} \dfrac{}{\gamma'; \delta_0 \vdash \psi_1(d')} \quad \text{init} \dfrac{}{\gamma_0, \psi_2; \delta_0 \vdash \psi_2}}{\gamma', (\psi_1(d') \to \psi_2); \delta_0 \vdash \psi_2}}{\gamma', \forall \psi; \delta_0 \vdash \psi_2}}{\gamma', \text{says}(a, \phi, b); \delta_0 \vdash \psi_2}}{\gamma', comm(a, b, \phi); \delta_0 \vdash \psi_2}$$

## A.2 Example: Fulfilling use-once obligations

Finally let's see a policy with !imp in it. The drink-beer example is fine. Suppose the bartender, $a$, has communicated the policy $\phi$ to a customer, $b$. The customer has to prove $po(drunk(b, \text{beer})) = drink(b, \text{beer})$. The abbreviations are:

$$\phi = \forall \psi \qquad \psi(x) = (\psi_1(x) \text{ !imp } \psi_2(\mathbf{x}))$$
$$\psi_1(x) = paid(x, 5\$) \qquad \psi_2(x) = \text{drink}(x, \text{beer})$$
$$\xi = comm(a, b, \phi) \qquad \xi' = paid(b, 5\$)$$

Now agent $b$ uses the evidence of the communication and the obligation $paid(b, 5\$)$ to prove that he is allowed to drink one beer having paid 5\$.

$$\cfrac{\cfrac{\cfrac{\cfrac{\cfrac{}{\gamma_0, \psi(b); \delta_0 \vdash (\psi_1(b) \text{ !imp } \psi_2(\mathbf{b}))} \text{ init}}{\gamma_0, \psi(b); \delta_0, \xi' \vdash \psi_2(b)} \text{ !imp\_l}}{\gamma_0, \forall \psi; \delta_0, \xi' \vdash \psi_2(b)} \text{ forall\_l}}{\gamma_0, \text{says}(a, \forall \psi, b); \delta_0, \xi' \vdash \psi_2(b)} \text{ SAY}}{\gamma_0, \xi; \delta_0, \xi' \vdash \psi_2(b)} \text{ OBS\_ACT}$$

## B Twelf Code

```
%% Object logic types
tp: type.
tm: tp -> type.

%% Objects in the Audit logic
agent: tp.
data: tp.
policy: tp.
action: tp.

%% Atomic Predicates
says: tm agent -> tm policy -> tm agent -> tm policy.
owns: tm agent -> tm data -> tm policy.
%% -- application-specific atomic predicates go here
print: tm agent -> tm data -> tm policy.

%% Policy grammar (with infix-operators for sugared writing)
imp: tm policy -> tm policy -> tm policy. %infix right 11 imp.
forall: (tm T -> tm policy) -> tm policy.
and: tm policy -> tm policy -> tm policy. %infix right 12 and.
%% -- note: only quantification over agents and data have introduction rules

%% Obligation implications for use-once !imp and use-many ?imp
?imp: tm action -> tm policy -> tm policy. %infix right 11 ?imp.
!imp: tm action -> tm policy -> tm policy. %infix right 11 !imp.

%% Core actions comm and creates
comm: tm agent -> tm agent -> tm policy -> tm action.
creates: tm agent -> tm data -> tm action.
%% -- application-specific actions can go here
printed: tm agent -> tm data -> tm action.

%% Conclusion derivation function
concl: tm action -> tm agent -> tm policy -> type.
%% Core concl relations
```

```
concl_comm: concl (comm A B Phi) B (says A Phi B).
concl_creates: concl (creates A Data) A (owns A Data).

%% Just 3 more objects: list, lin and nonlin
list: tp -> type. %%sequents are lists of lin and nonlin
nil: list X.
cons: tm X -> list X -> list X.
in: tm X -> list X -> type.
in_cons_head: in X (cons X Lx).
in_cons_tail: in X (cons Y Ly) <- in X Ly.
append: list X -> list X -> list X -> type.
append_nil: append nil X X.
append_nil2: append X nil X.
append_cons: (append (cons X XS) YS (cons X ZS)) <- (append XS YS ZS).

nonlin: tp. %% nonlinear for the unrestricted sequent Gamma
lin: tp.   %% linear for the linear sequent Delta

pol_c: tm policy -> tm nonlin. %% three types go into the nonlinear sequent
act_c: tm action -> tm nonlin.
?act_c: tm action -> tm nonlin.
!act_c: tm action -> tm lin. %% only actions in the linear sequent

%% The Entailment for the double context sequent calculus
entail: tm agent -> list nonlin -> list lin -> tm policy -> type.
%% -- (entail A Gamma Delta Phi) is in latex: \Gamma;\Delta \vdash_A \Phi

%% The 4 Audit logic "pivotal" rules
%% - - SAY
say: entail B (cons (pol_c Phi) Gamma) Delta Psi ->
       entail B (cons (pol_c (says A Phi B)) Gamma) Delta Psi.

%% - - OBS_ACT
obs_act: concl Act A Phi -> entail A (cons (pol_c Phi) Gamma) Delta Psi
                -> entail A (cons (act_c Act) Gamma) Delta Psi.
%% - - REFINE
refine:  entail A (cons (pol_c Phi) nil) nil Psi ->
        entail A (cons (pol_c (says A Phi B)) Gamma) Delta (says A Psi B).

%% - - DER_POL
%% -- a new relation, op, is needed for the der_pol rule
op: list nonlin -> tm policy -> type.
op_imp: (op Gamma Phi) -> (op Gamma (Psi imp Phi)).
op_forall: ({X: tm data} op Gamma (Phi X)) -> (op Gamma (forall Phi)).
op_says: (op Gamma Phi) -> (op Gamma (says B Phi C)).
op_owns: (op (cons (pol_c (owns A D)) Gamma) (owns B D)).
%% -- application specific predicates go here
op_print: (op (cons (pol_c (owns A D)) Gamma) (print B D)).

der_pol: (op Gamma Phi) ->
           entail A Gamma Delta Phi.

%% Common Inference rules for the Sequent Calculus
init: entail A (cons (pol_c Phi) Gamma) Delta Phi.
cut: entail A Gamma Delta Phi ->
     entail A (cons (pol_c Phi) Sigma) Epsilon Chi ->
     append Gamma Sigma Pi ->
     append Delta Epsilon Rho ->
```

```
        entail A Pi Rho Chi.
imp_l: entail A Gamma Delta Phi ->
       entail A (cons (pol_c Chi) Sigma) Epsilon Psi ->
       append Gamma Sigma Pi ->
       append Delta Epsilon Rho ->
         entail A (cons (pol_c (Phi imp Chi)) Pi) Rho Psi.
imp_r: entail A (cons (pol_c Phi) Gamma) Delta Chi ->
         entail A Gamma Delta (Phi imp Chi).
and_l1: entail A (cons (pol_c Phi) Gamma) Delta Chi ->
          entail A (cons (pol_c (Phi and Psi)) Gamma) Delta Chi.
and_l2: entail A (cons (pol_c Psi) Gamma) Delta Chi ->
          entail A (cons (pol_c (Phi and Psi)) Gamma) Delta Chi.
and_r: entail A Gamma Delta Phi ->
       entail A Sigma Epsilon Chi ->
       append Gamma Sigma Pi ->
       append Delta Epsilon Rho ->
         entail A Pi Rho (Phi and Chi).
%% -- as said only instances for agents and data
forall_l_agent: {X:tm agent} entail A (cons (pol_c (Phi X)) Gamma) Delta Chi ->
            entail A (cons (pol_c (forall Phi)) Gamma) Delta Chi.
forall_r_agent: ({X:tm agent} entail A Gamma Delta (Phi X)) ->
            entail A Gamma Delta (forall Phi).
forall_l_data: {X:tm data} entail A (cons (pol_c (Phi X)) Gamma) Delta Chi ->
            entail A (cons (pol_c (forall Phi)) Gamma) Delta Chi.
forall_r_data: ({X:tm data} entail A Gamma Delta (Phi X)) ->
            entail A Gamma Delta (forall Phi).
%% Obligation implications
!imp_l: entail A Gamma Delta ( Act !imp Psi) ->
              entail A Gamma (cons (!act_c Act) Delta) Psi.
!imp_r: entail A Gamma (cons (!act_c Act) Delta) Psi ->
            entail A Gamma Delta (Act !imp Psi).
?imp_l: entail A Gamma Delta (Act ?imp Psi) ->
              entail A (cons (?act_c Act) Gamma) Delta Psi.
?imp_r: entail A (cons (?act_c Act) Gamma) Delta Psi ->
            entail A Gamma Delta (Act !imp Psi).

%% Common Structural rules
w_l: entail A Gamma Delta Phi ->
       entail A (cons Chi Gamma) Delta Phi.
w_l_act: entail A Gamma Delta Phi ->
         entail A Gamma (cons Chi Delta) Phi.
contr_l: entail A (cons Phi (cons Phi Gamma)) Delta Chi ->
           entail A (cons Phi Gamma) Delta Chi.
perm_l: entail A Pi Delta Psi ->
        append Sigma (cons Phi (cons Chi Gamma)) Pi ->
        append Sigma (cons Chi (cons Phi Gamma)) Rho ->
          entail A Rho Delta Psi.
perm_act: entail A Gamma Delta Psi ->
          append Sigma (cons Act1 (cons Act2 Tau)) Delta ->
          append Sigma (cons Act2 (cons Act1 Tau)) Epsilon ->
            entail A Gamma Epsilon Psi.

%%%%%%%%%%%%%%%%%%%%%%%%%%%%
%% Formalization of example 2
%%%%%%%%%%%%%%%%%%%%%%%%%%%%
a: tm agent.
b: tm agent.
d: tm data.
```

```
rel: tm data -> tm data -> tm policy.
phi = forall [x:tm data]((rel d x) imp (print b d)).

ex2:entail a (cons (act_c (creates a d)) nil) nil (says a phi b)=
        obs_act concl_creates
            (cut
                (der_pol (op_says op_print))
                (refine
                    (forall_r_data [x] (imp_r (w_l init))))
                append_nil2 append_nil).

%% same account without the cut-rule
ex2b:entail a (cons (act_c (creates a d)) nil) nil (says a phi b)=
        obs_act concl_creates
            (der_pol (op_says (op_forall [x](op_imp op_print)))).
```